\documentclass[12pt]{article}

\catcode`@=12
\begin{document}
\thispagestyle{empty}
\begin{flushright} 
UCRHEP-T333\\ 
March 2002\
\end{flushright}
\vspace{0.5in}
\begin{center}
{\LARGE	\bf Quark Mass Matrices in the A$_4$ Model\\}
\vspace{1.5in}
{\bf Ernest Ma\\}
\vspace{0.2in}
{\sl Physics Department, University of California, Riverside, 
California 92521\\}
\vspace{1.5in}
\end{center}
\begin{abstract}\
If the standard model of quark interactions is supplemented by a discrete 
$A_4$ symmetry (which may be relevant for the lepton sector), the spontaneous 
breaking of the electroweak gauge symmetry allows arbitrary quark masses, but 
all mixing angles are predicted to be zero.  A pattern of the explicit 
breaking of $A_4$ is proposed, which results in a realistic charged-current 
mixing matrix.
\end{abstract}

\newpage
\baselineskip 24pt

In the standard model of quark interactions, the mass matrices of the $up$ 
and $down$ sectors are diagonalized separately, with unitary matrices 
$V_u$ and $V_d$ for $(u,c,t)_L$ and $(d,s,b)_L$ respectively.  The observed 
charged-current mixing matrix is then given by
\begin{equation}
V_{CKM} = V_u^\dagger V_d.
\end{equation}
There are two important issues to be understood here. (I) Why are quark masses 
hierarchical in each sector?  and (II) why is $V_{CKM}$ almost the identity 
matrix?  In this short note, there is no proposed answer to (I), but given 
(I), there is a possible explanation of (II) in the context of a recently 
proposed model \cite{mara} of nearly degenerate Majorana neutrino masses, 
using the discrete symmetry $A_4$.

There are 4 irreducible representations of $A_4$, i.e. \underline {1}, 
\underline {1}$'$, \underline {1}$''$, and \underline {3}, with the 
decomposition
\begin{equation}
\underline {3} \times \underline {3} = \underline {1} + \underline {1}' + 
\underline {1}'' + \underline {3} + \underline {3}.
\end{equation}
In particular,
\begin{eqnarray}
\underline {1} &=& a_1 a_2 + b_1 b_2 + c_1 c_2, \\ 
\underline {1}' &=& a_1 a_2 + \omega^2 b_1 b_2 + \omega c_1 c_2, \\ 
\underline {1}'' &=& a_1 a_2 + \omega b_1 b_2 + \omega^2 c_1 c_2, 
\end{eqnarray}
where the components of \underline {3} are denoted by $(a,b,c)$ and the 
complex number $\omega$ is the cube root of unity, i.e. $e^{2 \pi i/3}$. 
Hence $1 + \omega + \omega^2 = 0$.

Under $A_4$, the quarks are assumed to transform as follows.
\begin{eqnarray}
(u_i, d_i)_L &\sim& \underline {3}, \\ 
u_{1R}, ~d_{1R} &\sim& \underline {1}, \\ 
u_{2R}, ~d_{2R} &\sim& \underline {1}', \\ 
u_{3R}, ~d_{3R} &\sim& \underline {1}'',
\end{eqnarray}
in exact analogy with the left-handed lepton doublets and the right-handed 
charged-lepton singlets as proposed previously \cite{mara}.  The same three 
Higgs scalar doublets
\begin{equation}
\Phi_i = (\phi_i^+,\phi_i^0) \sim \underline {3}
\end{equation}
are also used.  Consequently, the Lagrangian of this model contains 
the following invariant Yukawa terms:
\begin{equation}
{\cal L}_Y = h^u_{ijk} \overline {(u_i,d_i)}_L u_{jR} \tilde \Phi_k + 
h^d_{ijk} \overline {(u_i,d_i)}_L d_{jR} \Phi_k + H.c.,
\end{equation}
where $\tilde \Phi_k = (\overline {\phi^0_k}, -\phi^-_k)$, and
\begin{eqnarray}
h^{u,d}_{i1k} &=& h^{u,d}_1 ~\delta_{ik}, \\ 
h^{u,d}_{i2k} &=& h^{u,d}_2 ~\delta_{ik} ~\omega^{i-1}, \\ 
h^{u,d}_{i3k} &=& h^{u,d}_3 ~\delta_{ik} ~\omega^{1-i}.
\end{eqnarray}

As $\phi_i^0$ acquire nonzero vacuum expectation values $v_i$, the quark mass 
matrices are of the form
\begin{equation}
{\cal M}_{u,d} = \left[ \begin{array} {c@{\quad}c@{\quad}c} h_1^{u,d} v_1 & 
h_2^{u,d} v_1 & h_3^{u,d} v_1 \\ h_1^{u,d} v_2 & h_2^{u,d} \omega v_2 & 
h_3^{u,d} \omega^2 v_2 \\ h_1^{u,d} v_3 & h_2^{u,d} \omega^2 v_3 & 
h_3^{u,d} \omega v_3 \end{array} \right].
\end{equation}
If the Higgs potential is invariant under $A_4$, it has been shown \cite{mara} 
that $v_1=v_2=v_3=v$ is a possible solution.  In that case, ${\cal M}_{u,d}$ 
is easily diagonalized, i.e.
\begin{equation}
{1 \over \sqrt 3} \left[ \begin{array} {c@{\quad}c@{\quad}c} 1 & 1 & 1 \\ 
1 & \omega & \omega^2 \\ 1 & \omega^2 & \omega \end{array} \right] 
{\cal M}_{u,d} = \left[ \begin{array} {c@{\quad}c@{\quad}c} \sqrt 3 h_1^{u,d} 
v & 0 & 0 \\ 0 & \sqrt 3 h_2^{u,d} v & 0 \\ 0 & 0 & \sqrt 3 h_3^{u,d} v 
\end{array} \right].
\end{equation}
This allows quark (and charged-lepton) masses to be hierarchical, even though 
neutrino masses are degenerate \cite{mara}.  Since both ${\cal M}_u$ and 
${\cal M}_d$ are diagonalized by the same unitary matrix, this model predicts 
$V_{CKM} = 1$ at this level, which is a good answer to (II).

\newpage
To obtain a realistic $V_{CKM}$, the $A_4$ symmetry must be broken.  Here 
the simple assumption is to add terms in ${\cal L}_Y$ of Eq.~(11) which are 
not just \underline {1} under $A_4$, but also \underline {1}$'$ and 
\underline {1}$''$, such that
\begin{equation}
|h''_i| << |h'_i| << |h_i|
\end{equation}
for each $i$.  In that case, the right-hand side of Eq.~(16) becomes 
proportional to
\begin{equation}
\left[ \begin{array} {c@{\quad}c@{\quad}c} h_1 & h'_2 & h''_3 \\ h''_1 & h_2 
& h'_3 \\ h'_1 & h''_2 & h_3 \end{array} \right],
\end{equation}
where the superscript $(u,d)$ has been dropped for simplicity.  Note also 
that $|h_1| << |h_2| << |h_3|$ in each sector because they are proportional 
to $(m_u, m_c, m_t)$ or $(m_d, m_s, m_b)$.  By rotating $q_{iR}$, the above 
matrix may be written as
\begin{equation}
\left[ \begin{array} {c@{\quad}c@{\quad}c} h_1 & h'_2 & h''_3 \\ 0 & h_2 & 
h'_3 \\ 0 & 0 & h_3 \end{array} \right]
\end{equation}
to a very good approximation, because $|h''_1| << |h_2|$ or $|h'_3|$, and 
$|h'_1|, |h''_2| << |h_3|$.  As a result, $V_{CKM}$ differs from the identity 
matrix by small amounts, i.e. \cite{ma01}
\begin{equation}
V_{us} \simeq {h'_2 \over h_2}, ~~ V_{ub} \simeq {h''_3 \over h_3}, ~~ 
V_{cb} \simeq {h'_3 \over h_3}.
\end{equation}
This explains why each is small, as well as why $|V_{ub}| << |V_{cb}|$.

Consider now the 3 Higgs doublets of this model.  Let
\begin{equation}
\left[ \begin{array} {c} \Phi \\ \Phi' \\ \Phi'' \end{array} \right] = 
{1 \over \sqrt 3} \left[ \begin{array} {c@{\quad}c@{\quad}c} 1 & 1 & 1 \\ 
1 & \omega & \omega^2 \\ 1 & \omega^2 & \omega \end{array} \right] 
\left[ \begin{array} {c} \Phi_1 \\ \Phi_2 \\ \Phi_3 \end{array} \right],
\end{equation}
then $\Phi$ has the properties of its standard-model counterpart, whereas 
$\Phi'$ and $\Phi''$ are degenerate in mass \cite{mara}.  Since only $\Phi$ 
has a nonzero vacuum expectation value, flavor-changing neutral currents are 
absent at tree level as far as $\Phi$ is concerned.  However, $\Phi'$ and 
$\Phi''$ have the following $predicted$ Yukawa interactions:
\begin{eqnarray}
{\cal L}_{int} &=& \left( {m_t \over v} \overline {(u,d)}_L t_R + 
{m_c \over v} \overline {(t,b)}_L c_R + {m_u \over v} \overline {(c,s)}_L
u_R \right) \tilde \Phi' \nonumber \\ &+& \left( {m_t \over v} \overline 
{(c,s)}_L t_R + {m_c \over v} \overline {(u,d)}_L c_R + {m_u \over v} 
\overline {(t,b)}_L u_R \right) \tilde \Phi'' \nonumber \\ &+& \left( 
{m_b \over v} \overline {(u,d)}_L b_R + {m_s \over v} \overline {(t,b)}_L 
s_R + {m_d \over v} \overline {(c,s)}_L d_R \right) \Phi' \nonumber \\ 
&+& \left( {m_b \over v} \overline {(c,s)}_L b_R + {m_s \over v} \overline 
{(u,d)}_L s_R + {m_d \over v} \overline {(t,b)}_L d_R \right) \Phi'' + H.c.
\end{eqnarray}
This shows that flavor-changing neutral currents involving only the first 
2 families are suppressed.  In fact, the most severe constraint comes from 
$B^0 - \overline {B^0}$ mixing which occurs through $(\phi')^0$ exchange:
\begin{equation}
{\Delta m_{B^0} \over m_{B^0}} \simeq {G_F m_b^2 \over 4 \sqrt 2} B_B f_B^2 
\left( {1 \over m_R^2} - {1 \over m_I^2} \right),
\end{equation}
where $G_F/\sqrt 2 = 1/12v^2$ and $m_{R,I}$ are the masses of the real and 
imaginary parts of $(\phi')^0$.  Using $f_B = 170$ MeV, $B_B = 1.0$, $m_b 
= 4.2$ GeV, and the experimental value \cite{pdg} of $5.9 \times 
10^{-14}$ for the above fraction, the condition
\begin{equation}
(m_R^{-2} - m_I^{-2})^{-1/2} >> 4.22 ~{\rm TeV}
\end{equation}
is obtained.  This means that $m_R$ and $m_I$ should be almost equal, if 
each is of order a few hundred GeV.

In conclusion, it has been shown how a realistic $V_{CKM} \simeq 1$ may be 
obtained in the context of the $A_4$ model of nearly degenerate Majorana 
neutrino masses.  It has specific verifiable predictions as given by Eq.~(22). 
In particular, the model requires $(\phi')^\pm$, $(\phi'')^\pm$ to have 
the same mass, Re$(\phi')^0$, Re$(\phi'')^0$ to have the same mass, and 
Im$(\phi')^0$, Im$(\phi'')^0$ to have the same mass.  Phenomenologically, 
the latter two pairs should also have almost the same mass, assuming that 
each is of order a few hundred GeV.

This work was supported in part by the U.~S.~Department of Energy under 
Grant No.~DE-FG03-94ER40837.

\newpage
\bibliographystyle{unsrt}

\end{document}